\documentclass[fleqn,10pt]{wlscirep}
\usepackage{graphicx}
\usepackage{mediabb}
\usepackage{amsmath,amssymb}
\title{Breaking limitation of quantum annealer in solving optimization problems under constraints}
\author[1,2,3*]{Masayuki Ohzeki}
\affil[1]{Graduate School of Information Science, Tohoku University, Sendai, Japan}
\affil[2]{Institute of Innovative Research, Tokyo Institute of Technology, Kanagawa, Japan}
\affil[3]{Sigma-i Co. Ltd., Tokyo, Japan}
\affil[*]{mohzeki@tohoku.ac.jp}
\begin{abstract}
Quantum annealing is a generic solver for optimization problems that uses fictitious quantum fluctuation.
The most groundbreaking progress in the research field of quantum annealing is its hardware implementation, i.e., the so-called quantum annealer, using artificial spins.
However, the connectivity between the artificial spins is sparse and limited on a special network known as the chimera graph.
Several embedding techniques have been proposed, but the number of logical spins, which represents the optimization problems to be solved, is drastically reduced.
In particular, an optimization problem including fully or even partly connected spins suffers from low embeddable size on the chimera graph.
In the present study, we propose an alternative approach to solve a large-scale optimization problem on the chimera graph via a well-known method in statistical mechanics called the Hubbard-Stratonovich transformation or its variants.
The proposed method can be used to deal with a fully connected Ising model without embedding on the chimera graph and leads to nontrivial results of the optimization problem.
We tested the proposed method with a number of partition problems involving solving linear equations and the traffic flow optimization problem in Sendai and Kyoto cities in Japan.
\end{abstract}
\begin{document}
\flushbottom
\maketitle
\thispagestyle{empty}
\section*{Introduction}
Quantum annealing (QA) is a generic algorithm aimed at solving optimization problems by exploiting the quantum tunneling effect.
The scheme was originally proposed as an algorithm for numerical computation \cite{Kadowaki1998} inspired by simulated annealing (SA) \cite{Kirkpatrick1983} and exchange Monte-Carlo simulation \cite{Hukushima1996}.
Moreover, its experimental realization has been accomplished recently and attracted significant attention.
Quantum annealing has the advantage of solving an optimization problem formulated with discrete variables.
A well-known example is searching for the ground state of the spin-glass model, which corresponds to various types of optimization problems, such as the traveling salesman problem and satisfiability problem \cite{Monnason1997,Monnason1998,Mezard2009}.
In QA, we formulate a platform to solve the optimization problem, the Ising model, and implement it in the time-dependent Hamiltonian.
The Hamiltonian takes the form of the formulated Ising model at the final time.
The initial Hamiltonian is governed by the ``driver'' Hamiltonian only with quantum fluctuation.
The frequently used driver Hamiltonian consists of the transverse field, which generates the superposition of the up and down spins.
The first stage of QA is initialized in the trivial ground state of the driver Hamiltonian.
The quantum effect will be gradually turned off, and will end so that only the classical Hamiltonian with a nontrivial ground state remains.
When the transverse field changes sufficiently slowly, the quantum adiabatic theorem ensures that we can find the nontrivial ground state at the end of QA \cite{Suzuki2005,Morita2008,Ohzeki2011c}.
Numerous reports have stated that QA outperforms SA \cite{Santoro2002,Santoro2004,Baldassi2018}.
The performance possibly stems from the quantum tunneling effect penetrating the valley of the potential energy.
The protocol of QA is realized in an actual quantum device using contemporary technology, namely, the quantum annealer \cite{Dwave2010a,Dwave2010b,Dwave2010c,Dwave2014}.
The output from the current version of the quantum annealer is not always the spin configuration in the ground state, due to the limitation of the device and environmental effects \cite{Amin2015}.
Therefore, several protocols based on QA do not keep the system in the ground state following the condition on the adiabatic quantum computation. Rather, they employ a nonadiabatic counterpart \cite{Ohzeki2010a,Ohzeki2011,Ohzeki2011proc,Somma2012} and the thermal effect \cite{Kadowaki2019}.
The quantum annealer has been tested for numerous applications, such as portfolio optimization \cite{Rosenberg2016}, protein folding \cite{Perdomo2012}, the molecular similarity problem \cite{Hernandez2017}, computational biology \cite{Richard2018}, job-shop scheduling \cite{Venturelli2015}, traffic optimization \cite{Neukart2017}, election forecasting \cite{Henderson2018}, machine learning \cite{Crawford2016,Arai2018nn,Takahashi2018,Ohzeki2018NOLTA,Neukart2018,Khoshaman2018}, and automated guided vehicles in plants \cite{Ohzeki2019}.

In addition, studies on implementing the quantum annealer to solve various problems have been performed \cite{Arai2018nn,Takahashi2018,Ohzeki2018NOLTA,Okada2019,Okada2019Potts,Okada2019Potts2}. 
The potential of QA might be boosted by the nontrivial quantum fluctuation, referred to as the nonstoquastic Hamiltonian, for which efficient classical simulation is intractable \cite{Seki2012,Seki2015,Ohzeki2017,Arai2018dy,Okada2019XX}.

The current version of the quantum annealer, the D-Wave 2000Q, employs the chimera graph, on which physical qubits are set.
The connection between the physical qubits is sparse and limited on the chimera graph.
Several embedding techniques are thus proposed, but the number of logical qubits, which represent the optimization problems to be solved, is drastically reduced \cite{Okada2019}.
In particular, the optimization problem, when it is written in terms of the Ising model, including fully or even partly connected spins, suffers from the smallness of the embeddable size on the chimera graph.
This is one of the bottlenecks in using D-Wave 2000Q.
The problem will remain in the near future because the limitation of the connection stems due to the design of the quantum circuits, which are not yet flexible.

In the present study, we propose an alternative way to solve a large-scale optimization problem with fully connected interactions between the logical qubits without any division into small subproblems and without embedding.
In statistical mechanics, a well-known traditional technique to tackle the fully connected interactions by hand is the Hubbard-Stratonovich transformation \cite{St1957,Hb1959} or its variants.
This technique mitigates the difficulty in dealing with the fully connected interaction emerging from the squared term by changing it into a linear term via the inversion of the Gaussian integral.
Using this technique, we formulate the original optimization problem with the squared term into another optimization problem with an equivalent linear term, but its coefficient can fluctuate stochastically.
To determine the value of the coefficient, we need to estimate the expectation value conditioned on that in the previous step.
In order to calculate the conditioned expectation values, we can utilize D-Wave 2000Q, which quickly outputs many samples of the Ising-variable configurations.
Therefore, we construct an iterative technique involving the following two processes: 1) determination of the coefficient and 2) estimation of the expectation values via D-Wave 2000Q instead of directly solving the original optimization problem.
By using our method, we can deal with a large-scale optimization problem even with the limited connections between the physical qubits, such as in D-Wave 2000Q.
In addition, our technique is not restricted to the case with the quantum annealer, such as the D-Wave 2000Q.
Complementary metal oxide semiconductor (CMOS) annealing, which also has the same bottleneck \cite{Yamaoka2016}, is within the range of application.
Furthermore, our technique provides an alternative method to formulate the combinatorial optimization problem with several constraints.
In this sense, even if the full connectivity of the logical variables in the hardware is realized such as the Fujitsu digital annealer \cite{Tsukamoto2017}, our technique is valuable for finding optimal solutions.
The listed hardware are special-purpose ones invented to solve the Ising model to find the minimizer of the cost function written in its form and quickly attain the sampling following particular distribution functions.

Our technique is closely related to our previous study on adaptive quantum Monte-Carlo simulation \cite{Ohzeki2017}.
In the previous study, we showed that the fully connected antiferromagnetic interaction, which results in the sign problem via naive classical-quantum mapping, such as the Suzuki-Trotter decomposition \cite{Suzuki1976}, can be transformed into a fluctuating transverse field without any sign problems, by using the Hubbard-Stratonovich transformation.
In this case, one needs to estimate the expectation value of the transverse magnetization by the quantum Monte-Carlo simulation or message-passing algorithm as an approximate way \cite{Ohzeki2019JPSJ}.
A similar approach is found in the simulation for the strongly correlated electrons \cite{White1989}.

The remaining part of the present paper consists of the following contents.
In the next section, we show how to transform the optimization problem with squared terms into an equivalent simplified model and describe our method.
In the following sections, we test our method with the number partition problem, which is a typical problem with a fully connected interaction, and solving an inverse problem from a small number of equations, which is also the case.
In addition, several results for the optimization problem under several constraints are demonstrated.
In the last section, we summarize our study.

\section*{Problem Setting}
In QA, we formulate the optimization problem as the target Hamiltonian.
In addition to the target Hamiltonian, we employ the driver Hamiltonian, which generates the quantum fluctuation driving the system.
We consider the case with the target Hamiltonian $f({\bf q})$ where ${\bf q} = (q_1,q_2,\cdots,q_N)$, and $q_i$ is a binary variable as $0$ and $1$.
The binary variable can be written as the $z$-component of the Pauli matrices, which represent the logical qubits, namely, the Ising variables as $q_i = (1+\sigma_i)/2$.
Against the longitudinal Ising spin variables, the induction of the transverse field generates superposition to search the ground state efficiently and solves various optimization problems.
In practical use of QA, several sets of the variables must satisfy the constraints.
As is often the case, squared terms expressing the constraints appear.
This is called the penalty method in the context of the optimization problems.
We assume here that the target Hamiltonian consists of several summations of the squared terms representing the constraints such as $F_i({\bf q}) = C_i~\forall i$ and the other terms $f_0({\bf q})$ as
\begin{equation}
f({\bf q}) = f_0({\bf q}) + \frac{1}{2}\sum_{i}\lambda_i \left( F_i({\bf q}) - C_i \right)^2,
\end{equation}
where $\lambda_i$ is a predetermined parameter set to be relatively large because the squared terms often express the constraints for the Ising variables.
The squared terms yield fully connected interactions among the Ising variables because $F_i({\bf q})$ often consists of the summation over several elements of ${\bf q}$.

We introduce several examples appearing in QA.
In a practical application of the quantum annealer for the reduction of traffic flow \cite{Neukart2017}, the squared term is then employed as
\begin{equation}
f({\bf q}) - f_0({\bf q}) = \frac{\lambda}{2}\sum_{i}  \left( \sum_{\mu} q_{\mu,i} -1\right)^2\label{TF1},
\end{equation}
where 
\begin{equation}
f_0({\bf q}) = \frac{1}{2}\sum_e \left( \sum_{\mu} \sum_i S_{e,\mu, i}q_{\mu,i}\right)^2, \label{TF2}
\end{equation}
and $q_{\mu,i} = (1+\sigma_{\mu,i})/2$ represents the selection of the $\mu$-th route for the $i$-th car and $S_{e,\mu,i}$ denotes the occupation of the road segment $e$ by the $\mu$ th route and $i$ th car.
To avoid traffic congestion, they also implement another squared term as in the second term.

Furthermore, the cost function for inferring the $N$-dimensional original signal $\sigma_k$ from the output consisting of $M$ linear combinations $y_{\mu} = \sum_{k}a_{\mu k} q^0_k$ in wireless communications and signal processing is written as
\begin{equation}
f({\bf q}) = \frac{1}{2}\sum_{\mu=1}^M \left( y_{\mu} - \sum_{k=1}^N a_{\mu k} q_k\right)^2.\label{CDMA}
\end{equation}
One of the fascinating applications of QA is Q-Boost \cite{Neven2012}, which selects relevant weak classifiers to gain the performance by combining them.
In addition, the squared terms stem from the penalty method for the constraints as follows
\begin{equation}
f_0({\bf q}) = \frac{1}{2}\sum_{\mu} \left( y_{\mu} - \sum_{k}u_k({\bf x}_{\mu}) q_k\right)^2,\label{QBoost}
\end{equation}
where $u_k({\bf x}_{\mu})$ denotes the weak classifier, ${\bf x}_{\mu}$ is the data vector, and its label is $y_{\mu}$.
If the number of the classifiers is set to be $K$, the additional squared term is employed as
\begin{equation}
f({\bf q}) - f_0({\bf q}) = \frac{\lambda}{2} \left( \sum_{k=1}^N q_k - K \right)^2.
\label{QBoost+}
\end{equation}

In addition, the cost function itself is often expressed in the square form.
One of the examples is the number partition problem, given as
\begin{equation}
f({\bf q}) = f_0({\bf q}) = \frac{1}{2} \left( \sum_{i=1}^N n_i \sigma_i \right)^2, \label{NP}
\end{equation}
where $n_i$ denotes a component of the numbers to be divided into two groups.
One group is assigned $\sigma_i=+1$ and the other $\sigma_i = -1$.
The summation $\sum_i n_i \sigma_i$ is desired to be zero to divide the component into two equal groups with an equal summation of the numbers.

As exemplified above, to solve various optimization problems in the quantum annealer through the formulation of the quadratic unconstrained binary optimization (QUBO) problem, we often implement the squared terms.
However, the squared terms result in fully connected interactions among the Ising variables included in them.
The fully connected interactions prevent efficient computation in the current version of the quantum annealer because it has a limitation in dealing with the connection between artificial spins.
For example, we have to embed the original optimization problem with fully connected interactions into a sparse graph, known as the chimera graph, on the superconducting chip for the case of D-Wave 2000Q.
Then, unfortunately, although D-Wave 2000Q has over $2000$ active physical qubits, the number of logical qubits dealt with is reduced to about $60$ in the worst case.
This is one of the crucial bottlenecks of the current version of the quantum annealer.
A different type of special-purpose hardware implementing the Ising model, namely, CMOS annealing, also has the same bottleneck \cite{Yamaoka2016}, whereas the Fujitsu digital annealer is free from the problem in connectivity \cite{Tsukamoto2017}.
To avoid the difficulty of the squared terms, in the previous study, a different type of driver Hamiltonian from the transverse field was proposed \cite{Itay2016-1,Itay2016-2}.
In this case, they succeeded in enhancing the performance of QA, but the current version of the quantum annealer cannot employ their method.
Below, we mitigate this difficulty by tackling this problem in the optimization problem with squared terms using the standard method in statistical mechanics.
The method proposed below is available in the current version of the quantum annealer.

\section*{Reduction of squared terms}
A well-known technique for reducing the squared terms into linearized ones is the Hubbard-Stratonovich transformation or its variants \cite{St1957,Hb1959}.
First, we take the partition function expressing the equilibrium state governed by our target Hamiltonian as
\begin{equation}
Z = \sum_{\bf q} \exp\left( -\beta f({\bf q})\right).
\end{equation}
where $\beta$ is the inverse temperature.
Here, we perform the Hubbard-Stratonovich transformation of the squared terms and obtain another expression for the partition function as
\begin{equation}
Z = \sum_{\bf q} \prod_{k}\int Dz_k\exp\left( i\sum_{k}\sqrt{\beta \lambda_k} z_k\left( F_k({\bf q}) - C_k\right) -\beta f_0({\bf q})\right),
\end{equation}
where $\int Dz_k =\int dz_k \exp(-z_k^2/2)/\sqrt{2\pi}$.
Then, we change the integral variable $z_k \to - i \sqrt{\beta/\lambda_k} \nu_k$.
The resulting partition function is 
\begin{equation}
Z \propto \sum_{\bf q} \prod_{k}\int d\nu_k\exp\left( \sum_{k}\frac{\beta}{2\lambda_k} \nu_k^2  + \beta \sum_{k} \nu_k\left( F_k({\bf q}) - C_k\right) -\beta f_0({\bf q})\right).
\end{equation}
We obtain an effective Ising model with linear terms on the constraints and continuous variables ${\boldsymbol \nu}=(\nu_1,\nu_2,\cdots)$, namely the Lagrange multipliers.
The effective Hamiltonian is
\begin{equation}
H({\bf q},{\boldsymbol \nu}) = - \sum_{k} \frac{\nu^2_k}{2\lambda_k }- \sum_{k} \nu_k \left( F_k({\bf q}) - C_k\right) + f_0({\bf q}).
\end{equation}
The remaining problem is the minimization of the effective Hamiltonian instead of the original Hamiltonian.
This is the same technique for dealing with constraints in optimization problems such as the Lagrange multiplier method.
The original formulation employing squared terms is the penalty method.
In the penalty method, we have to take a relatively large value of the coefficients $\lambda_k$ to deal with the constraints.
However, the large value of the coefficients leads to obstacles in the optimization by the current version of the quantum annealer because it has the limitation of range and interval of the coefficient.
In our formulation, we can take the limit of $\lambda_k \to \infty$ in a straightforward way.
Instead of the large coefficient, the adaptive change of the multiplier $\nu_k$ retains the constraints. 
As a ``dual" problem, the effective Hamiltonian for ${\boldsymbol \nu}$ can be obtained as
\begin{equation}
H({\boldsymbol \nu}) = - \sum_{k} \frac{\nu^2_k}{2\lambda_k } + \sum_{k} \nu_k C_k - \frac{1}{\beta}\log Z({\boldsymbol \nu}),
\end{equation}
where $Z({\boldsymbol \nu})$ is the effective partition function defined as
\begin{equation}
Z({\boldsymbol \nu}) = \sum_{\bf q} \exp\left( -\beta f_0({\bf q}) + \beta \sum_{k} \nu_k  F_k({\bf q}) \right).
\end{equation}
Obviously, the effective Hamiltonian for ${\boldsymbol \nu}$ is highly nontrivial.
In other words, the complexity of the original optimization problem remains even in the dual problem with continuous variables.
The minimizer of the effective Hamiltonian is the saddle point of the integrand in the partition function when we take the limit of $\beta \to \infty$.
The saddle point equation is given as 
\begin{equation}
C_k - \left\langle F_{k}({\bf q})\right\rangle_{\bf q} = 0, 
\end{equation}
where the bracket denotes expectation by the probability distribution of ${\bf q}$ conditioned on the value of ${\boldsymbol \nu}$.
Notice that, in general, the free energy for the so called spin-glass models, as discussed in context of the optimization problem, has many local minima.
Thus, the saddle point is not unique.
This is a consequence of the non-monotonic increase in $\left\langle F_{k}({\bf q})\right\rangle_{\bf q}$ against ${\boldsymbol \nu}$ in the spin glass models.
We emphasize that the complexity to find the ground state remains even by our method.
In this sense, our method is strongly dependent on the form of $f_0({\bf q})$.

Our remaining problem is to attain the saddle point by gradually changing the value of ${\boldsymbol \nu}$.
To find the saddle point, one may utilize the steepest ascent method.
We take $\beta \to \infty$, and the expectation value is evaluated by the Ising spin configuration in the ground state.
Thus, the sampling of the spin configuration by use of the D-Wave 2000Q can be performed.
In particular, the practical optimization problem has a nontrivial cost function $f_0({\bf q})$.
Then, the computation of the expectation value as $\left\langle F_{k}({\bf q})\right\rangle_{\bf q}$ is harmful.
To mitigate its difficulty, the special-purpose machine is valuable.
Notice that our technique is not restricted to use of the D-Wave 2000Q.
Our technique is helpful for the Fujitsu digital annealer and CMOS annealing chip to enhance the precision to satisfy the constraints. 

Instead of the direct manipulation of the Hubbard-Stratonovich transformation, we may consider the variational free energy, namely the Gibbs free energy.
Let us consider the target Hamiltonian without squared terms for constraints, namely $f_0({\bf q})$.
Then, the Gibbs-Boltzmann distribution is given as $P({\bf q}) = \exp\left( -\beta f_0({\bf q})\right)/Z$.
We introduce the Kullback-Leibler divergence to measure the distance between the trial distribution function $P$ and $Q$ as
\begin{equation}
{\rm KL}(P|Q) = \sum_{\bf q} Q({\bf q}) \log \left( \frac{Q({\bf q})}{P({\bf q})} \right).
\end{equation}
Here, we consider the trial distribution $Q$ with minimum distance from $P$, whereas the expectation satisfies the following constraints
\begin{equation}
C_k = \left\langle F_k({\bf q}) \right\rangle_{Q}.
\end{equation}
The minimization of the KL divergence under this constraint yields the Gibbs free energy as
\begin{equation}
G({\bf C}) = \min_{Q}\left\{E[Q]-S[Q]|~C_k = \left\langle F_k({\bf q}) \right\rangle_{Q}~\forall k\right\},
\end{equation}
where $E[Q] = \sum_{\bf q}Q({\bf q})f_0({\bf q})$ and $S[Q] = -Q({\bf q})\log Q({\bf q})$.
Here, we introduce the Lagrange multiplier ${\boldsymbol \nu}$ for solving the minimization problem under the constraints.
The minimizer depending on the Lagrange multiplier ${\boldsymbol \nu}$ can be attained in a straightforward way as
\begin{equation}
Q({\bf q}) = \frac{1}{Z({\boldsymbol \nu})}\exp\left( - \beta f_0({\bf q}) + \sum_{k}\nu_k F_k({\bf q})\right).
\end{equation}
Then, the Gibbs free energy is written as
\begin{equation}
G({\bf C }) = \max_{\boldsymbol \nu}\left\{ \sum_{k}\nu_kC_k - \log Z({\boldsymbol \nu})\right\}.
\end{equation}
This corresponds to the minimization of the effective Hamiltonian for ${\boldsymbol \nu}$.
The Gibbs free energy is the starting point of the Plefka expansion to establish a systematic way to solve the Ising spin-glass model beyond the level of the mean-field theory.
Then, we consider the weak-interaction limit for computing the summation of the logarithmic term.
In contrast, we have an efficient sampler for estimating the expectation of the Ising spin-glass model such as the D-Wave 2000Q.
Therefore, we do not need any approximation to proceed our formulation further to solve the optimization problem under several constraints.
When $F_k({\bf q})$ consists of the summation over several binary variables, all we have to do is induce the longitudinal magnetic field to realize the resultant effective Hamiltonian.
Thus, we construct a simple algorithm to solve the optimization problem as follows.

\begin{itemize}
\item Initialize the Lagrange multipliers ${\boldsymbol \nu}^{t=0}$.
\item Compute the gradient and update the Lagrange multipliers as
\begin{equation}
\nu^{t+1}_k = \nu^{t}_k + \eta\left( C_k - \left\langle F_k({\bf q}) \right\rangle_{Q^t} \right),\label{update}
\end{equation}
where $\eta$ is a step width of the gradient method to achieve the maximization and $Q^t$ is the trial distribution function with ${\boldsymbol \nu}={\boldsymbol \nu}^t$.
\end{itemize}
Let us briefly describe the above procedure for the case with a constraint for the simple summation, namely $F_k({\bf q}) = \sum_{i}q_i$.
The initial condition, for instance, is set to be no biases on the system.
Then, $\langle F_k({\bf q}) \rangle_{Q^t}$ takes a finite value depending on $f_0({\bf q})$.
When $\langle F_k({\bf q}) \rangle_{Q^t} > C_k$, $\nu^{t}_k$ decreases for reducing $\langle F_k({\bf q}) \rangle_{Q^t}$ and vice versa.
The convergence depends on the rate of the update $\eta$.
One may utilize the line search for an optimal choice of $\eta$ to efficiently attain convergent behavior.
To solve the constraints, we need to estimate the expectation of the Ising spin glass with the Hamiltonian $f_0({\bf q}) - \sum_{k}\nu_k F_k({\bf q})$, in which the squared terms on the constraints are absent.
This is much easier to implement it in the D-Wave 2000Q and CMOS annealing chip with finite connectivity of the graph.
Notice that several optimization problems are written only by the squared terms, namely $f_0({\bf q})=0$.
Then, the effective Hamiltonian consists only of linear terms, namely, the local magnetic fields.
In this sense, it is not necessary to use the special-purpose device to generate the sampling of the nontrivial Hamiltonian such as the D-Wave 2000Q.
Our technique from this perspective would be quite valuable for the case with nontrivial $f_0({\bf q})$.

\section*{Experiments}
We test our method with various problems.
The first experiment is performed for selection of the $K$-minimum set of the $N$ random values.
The original cost function is written as 
\begin{equation}
f({\bf q}) = \sum_{i=1}^N h_i q_i + \frac{\lambda}{2}\left( \sum_{i=1}^N q_i - K\right)^2, \label{simple}
\end{equation}
where $h_i$ takes a random value following the uniform distribution.
We set $N=2000$ and $K=5$.
The square term in Eq. (\ref{simple}) often appears in application of the quantum annealer to the optimization problem under constraints.
The standard approach for solving optimization problems as in Eq. (\ref{simple}) using the D-Wave 2000Q is embedded on the chimera graph up to $64$ logical variables.
However, our technique can embed $2000$ logical variables directly.
In this case, because $f_0({\bf q})=0$, we do not necessarily need the sampling from the special-purpose devices. 
This is just a test for validation of our technique.
In addition, the first term in Eq. (\ref{simple}) is very simple but the exact solution is attained in a straightforward manner.
We can check the validity of our technique.
The initial condition is set as $\nu^0=0$.
We take the step width for the update in the steepest ascent by line search in all the cases shown below.

As shown in Fig. \ref{simple_ene}, we confirm that our technique can select $K$-minimum set from $N$ random variables and reach the optimal solutions.
We plot the residual energy, which is the difference between the cost function and its minimum value.
\begin{figure}
\begin{center}
\begin{tabular}{c}
\begin{minipage}{0.5\textwidth}
\includegraphics[width = 1.0\textwidth]{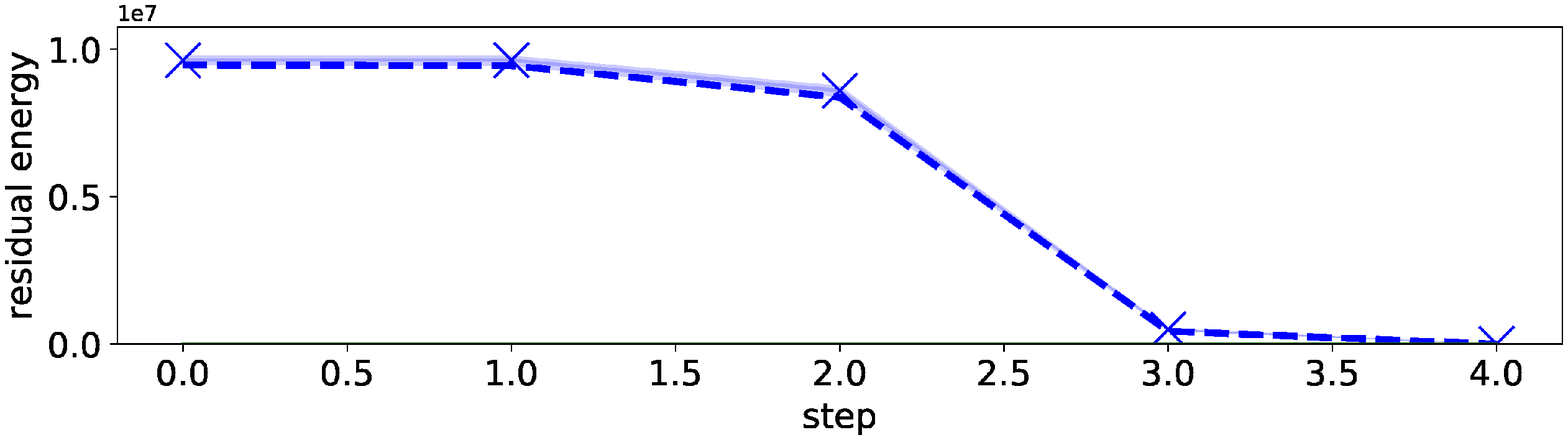}
\end{minipage}
\begin{minipage}{0.5\textwidth}
\includegraphics[width = 1.0\textwidth]{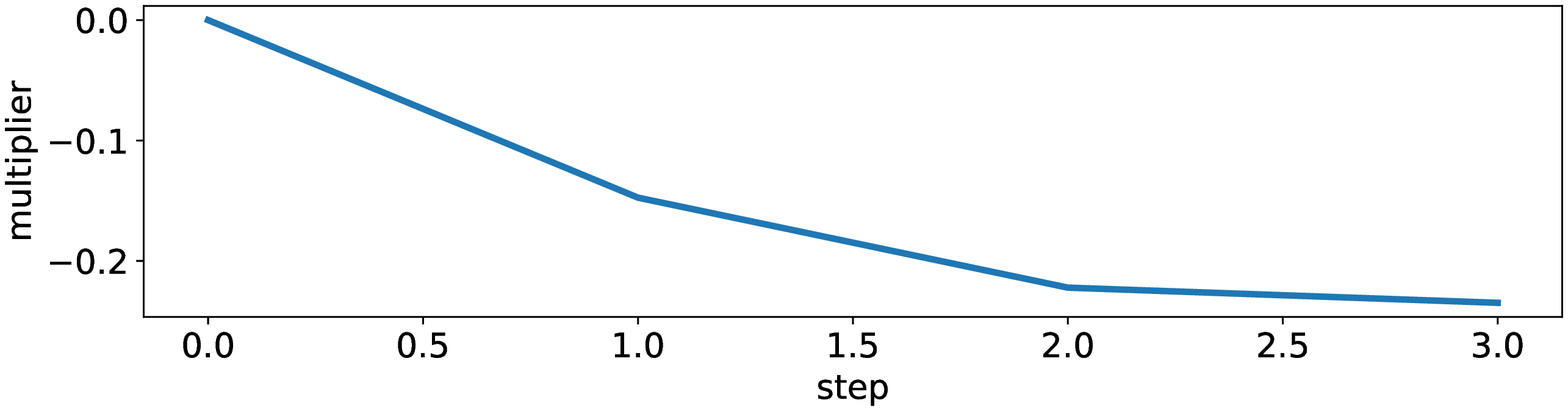}
\end{minipage}
\end{tabular}
\caption{Residual energy (left) and multiplier (right) at each step in our method for selection of the $K$-minimum set of the $N$ random variables.
In the upper panel, the cross points represent the empirical average of output from the D-Wave 2000Q, and the dashed curves denote the minimum value.
}
\label{simple_ene}
\end{center}
\end{figure}

The second experiment is performed on the number partition problem as in Eq. (\ref{NP}).
Then, the effective Hamiltonian is
\begin{equation}
H({\bf q},\nu^t) = \nu^t\sum_{i =1}^N n_{i} q_i. \label{np_eq}
\end{equation}
For the number partition problem, $f_0({\bf q})=0$.
Therefore, we do not need the sampling from the special-purpose device.
This is just a test for validation of our technique.
We set $2000$ components of integer numbers and permute them randomly because our available system of D-Wave 2000Q has about $2000$ active physical qubits.
The standard formulation of the number partition as in Eq. (\ref{NP}) by the D-Wave 2000Q is embedded on the chimera graph up to $64$ logical variables.
In contrast, our method, on the chip of D-Wave 2000Q, sets the $2000$ local magnetic field for each physical qubit without any embedding techniques.

\begin{figure}
\begin{center}
\begin{tabular}{c}
\begin{minipage}{0.5\textwidth}
\includegraphics[width = 1.0\textwidth]{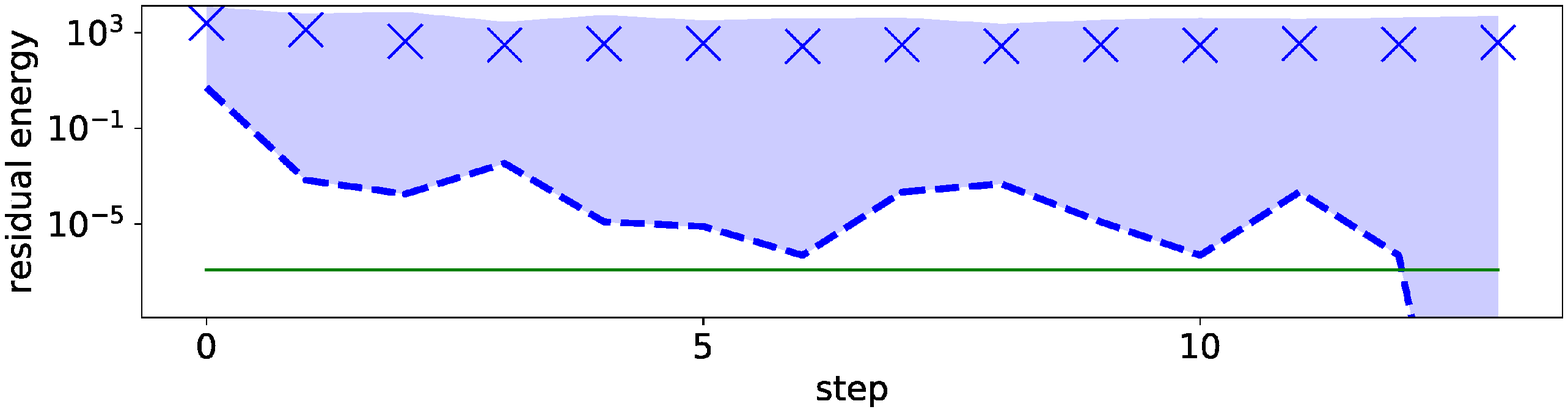}
\end{minipage}
\begin{minipage}{0.5\textwidth}
\includegraphics[width = 1.0\textwidth]{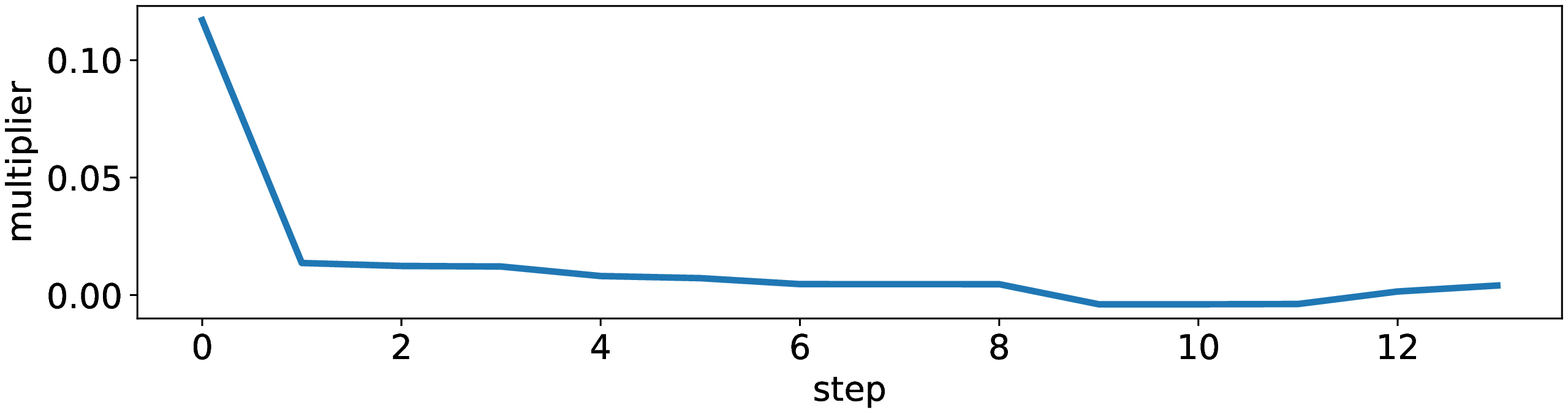}
\end{minipage}
\end{tabular}
\caption{Residual energy (left) and multiplier (right) at each step in our method for the number partition problem.
The cross points represent the empirical average of output from the D-Wave 2000Q, and the dashed curves denote the minimum value.
The green line denotes the minimum unit of the residual energy as $(1/N)^2/2$ in the number partition problem.
When the energy becomes lower than the green line, the solution reaches the optimal one.
}
\label{np_ene}
\end{center}
\end{figure}

In Fig. \ref{np_ene}, we plot the cost function (\ref{NP}) and not the value of the effective Hamiltonian (\ref{np_eq}) at each step of our method.
In addition, the lower figure shows the multiplier $\nu^t$.
We employ the steepest ascent to attain the saddle point.
The saddle point is expected to be around $\nu=0$ for the number partition problem because even tiny strength of the magnetic field enforces all the spin directions to be positive or negative but the optimal solution is randomly oriented.
The initial condition is set to be $\nu^0=0.2$.
The result confirms that the optimal solution can be obtained by our technique.

\begin{figure}
\begin{center}
\begin{tabular}{c}
\begin{minipage}{0.5\textwidth}
\includegraphics[width = 1.0\textwidth]{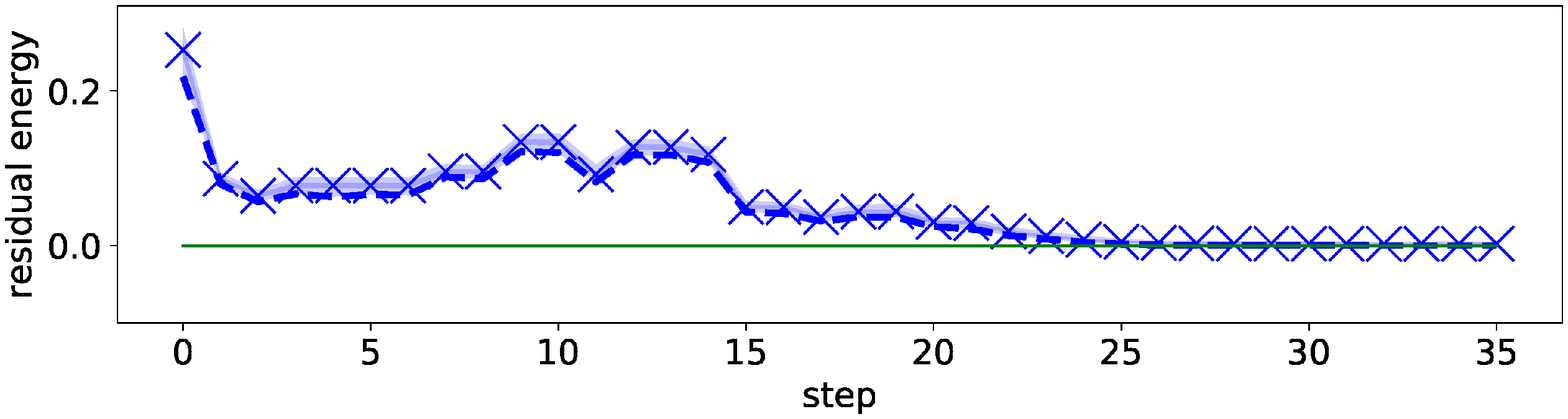}
\end{minipage}
\begin{minipage}{0.5\textwidth}
\includegraphics[width = 1.0\textwidth]{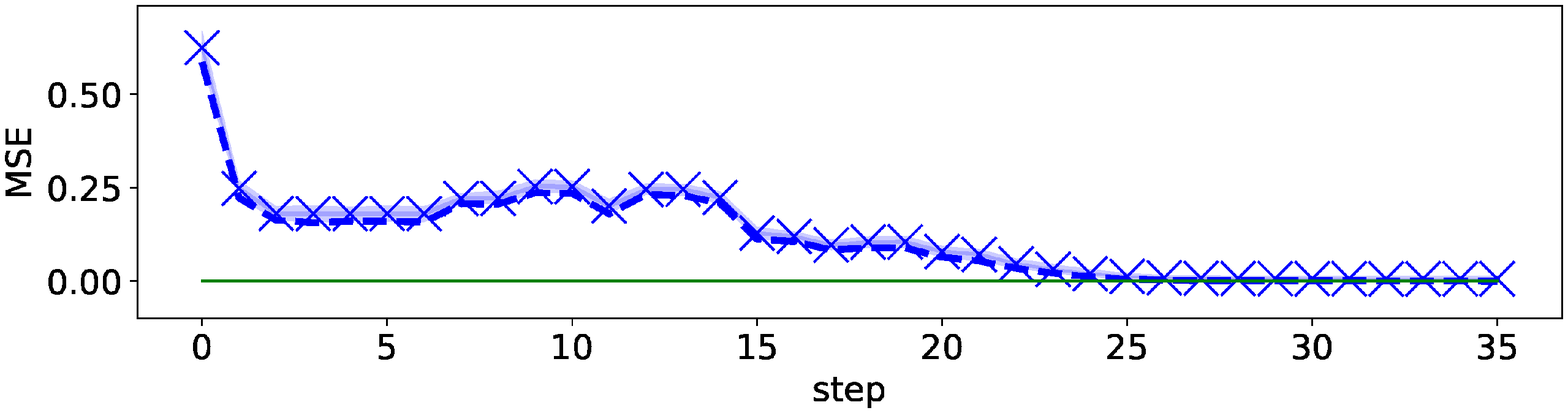}
\end{minipage}
\end{tabular}
\caption{Residual energy (left) and MSE (right) at each step in our method for solving the linear equation.
The same symbols are used as in Fig. \ref{np_ene} except for the green line.
In this figure, the green line denotes zero.
}
\label{cdma_ene}
\end{center}
\end{figure}

The third example is solving the linear equation.
In other words, it is termed as inference of the $N$-dimensional input from the $M$ linear combinations as in Eq. (\ref{CDMA}).
The effective Hamiltonian is
\begin{equation}
H({\bf q},{\boldsymbol \nu}) = {\boldsymbol \nu}^{\rm T} A {\bf q} 
\end{equation}
and $f_0({\bf q})=0$.
Again in this case, the effective Hamiltonian consists of only the local magnetic fields.
Thus, D-Wave 2000Q can solve the inference problem for over $2000$ dimensional inputs.
We prepare the linear combination of the original signal ${\bf q}^0$ as ${\bf y} = A{\bf q}^0$, where $A$ is a $M\times N$ matrix with random elements following the Gaussian distribution with a vanishing mean and unit variance, and $q^0_i = 0$ and $1$ follows an equal distribution. 
When $\alpha = M/N > 0.633$ obtained by analysis in statistical mechanics \cite{Tanaka2006}, the inference of the $N$-dimensional original signal from $M$ linear combinations by solving the optimization problem can be successful.
In other words, the solution ${\bf q}$ can coincide with the original signal ${\bf q}^0$.
We set $M/N = 0.8$ and $N=2000$.
As shown in Fig. \ref{cdma_ene}, we successfully find the perfect reconstruction of the original input.
We observe the residual energy and mean squared error (MSE) defined as $\left\| {\bf q} -{\bf q}_0 \right\|_2^2/N$.
When the MSE and residual energy become zero, the perfect reconstruction is realized.

\begin{figure}
\begin{center}
\includegraphics[width = 0.2\textwidth]{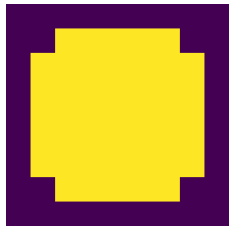}
\caption{Example of two-dimensional structure.
}
\label{example_2D}
\end{center}
\end{figure}

\begin{figure}
\begin{center}
\begin{tabular}{c}
\begin{minipage}{0.5\textwidth}
\includegraphics[width = 1.0\textwidth]{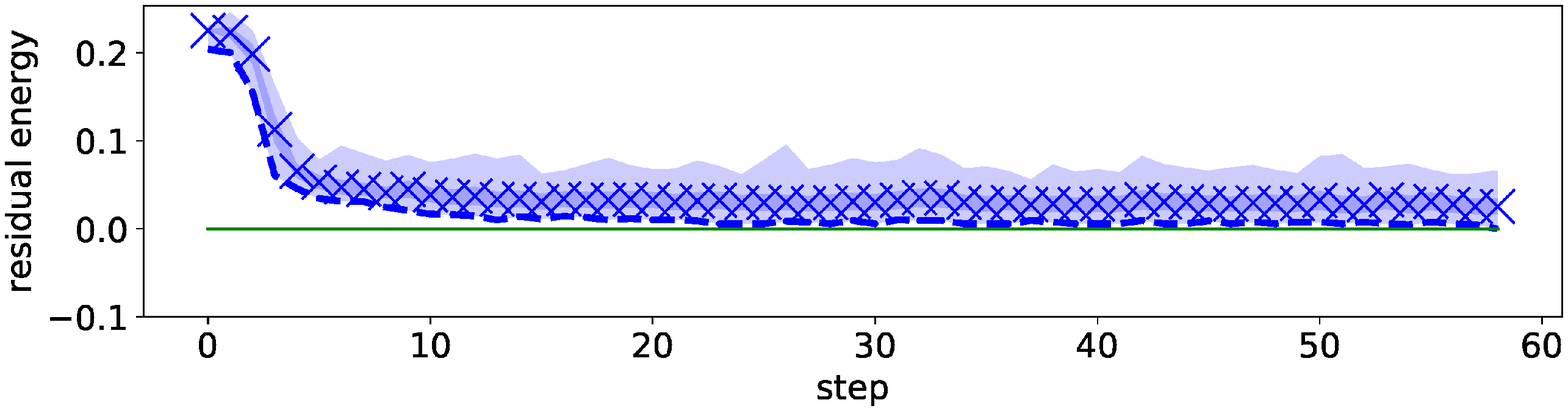}
\end{minipage}
\begin{minipage}{0.5\textwidth}
\includegraphics[width = 1.0\textwidth]{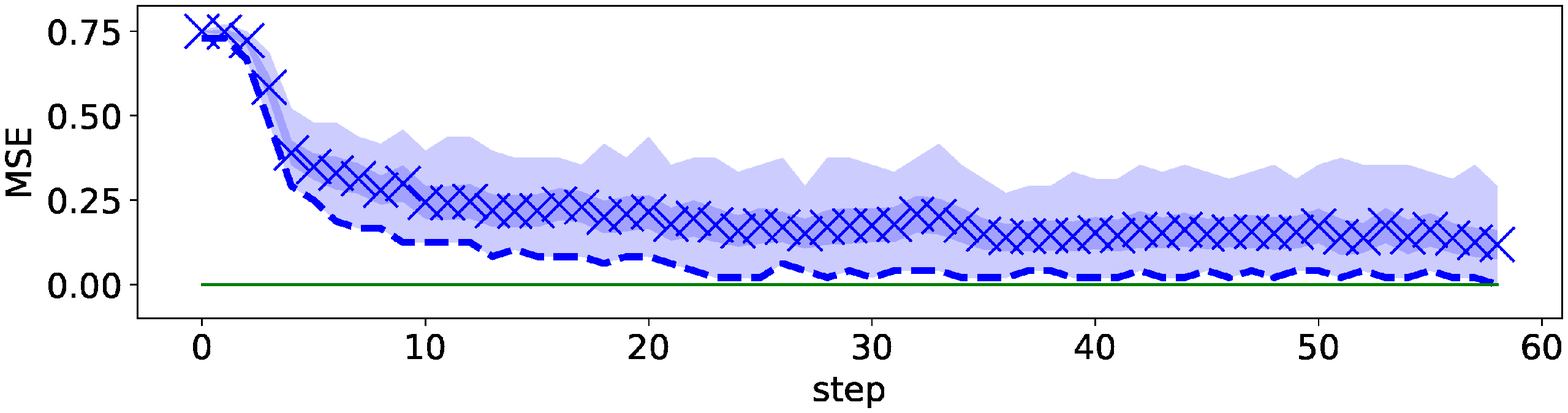}
\end{minipage}
\end{tabular}
\caption{Residual energy and MSE at each step in our method for inferring two-dimensional images.
The same symbols are used as in Fig. \ref{np_ene} except for the green line.
In this figure, the green line denotes zero.
}
\label{cdma2D}
\end{center}
\end{figure}

The fourth example is essentially the same problem as the previous one.
However, the original input represents the two-dimensional structure as shown in Fig. \ref{example_2D}
The problem emerges typically in the compressed sensing for reconstructing an original input from insufficient number of outputs using its sparsity as prior information.
Let us take an example like ${\bf q}_0$ in the previous case as in Fig. \ref{example_2D}, which is two-dimensional structured data, while all the non-zero components are connected to each other.
For $\alpha = 0.6$, in which the number of outputs is too small to recover the original input, we employ the following Hamiltonian to infer the original input 
\begin{equation}
f({\bf q}) = \sum_{\langle ij \rangle }q_i q_j + \frac{\lambda}{2}\sum_{\mu=1}^M \left( y_{\mu} - \sum_{k=1}^N a_{\mu k} q_k\right)^2. \label{CS}
\end{equation}
In this case, we regard the first term as $f_0({\bf q})$.
Then, sampling using the D-Wave 2000Q is efficient to evaluate the expectation value in the update equation (\ref{update}).
As shown in Fig. (\ref{cdma2D}), we demonstrate that our method solves the optimization problem written in Eq. (\ref{CS}) even for the insufficient outputs $M/N<0.633$.

The fifth example is the traffic flow optimization problem.
Following the previous study \cite{Neukart2017}, we extract the route data from the OpenStreetMap via osmnx \cite{Boeing2017}.
We prepare candidate routes for each car by the shortest path, and its variants. 
We then assign the binary variables $q_{\mu,i}$ for each car $i$ and its route $\mu$.
Each car selects a single route by satisfying the constraints as in Eq. (\ref{TF1}).
Instead of the constraints, we may decompose the quadratic term in $f_0({\bf q})$ as follows
\begin{equation}
\sum_{e} \left( \sum_{\mu,i} S_{\mu,i,e} q_{\mu,i} \right)^2\to \sum_{e} \nu_e \left(\sum_{\mu,i} S_{\mu,i,e}q_{\mu,i}\right).
\end{equation}
Then, the effective Hamiltonian can be reduced to the Ising model in the local-magnetic fields as
\begin{equation}
H({\bf q},{\boldsymbol \nu}) = - \sum_{\mu,i}h_{\mu,i} q_{\mu,i} + \frac{\lambda}{2}\sum_{i}\left( \sum_{\mu} q_{\mu,i} -1 \right)^2,
\end{equation}
where $h_{\mu,i} = \sum_{e}\nu_e S_{\mu,i,e}$.
Owing to the reduction of $f_0({\bf q})$ instead of the constraints, notice that we can easily attain the expected value of the effective Hamiltonian without any sampling method as follows
\begin{equation}
\langle q_{\mu,i} \rangle = \delta_{\mu=\mu^*_i},\label{exact}
\end{equation}
where $\mu^*_i = \arg \max_{\mu} \left(h_{\mu,i} \right)$.
Then, the updated equation for each ${\boldsymbol \nu}$ leads to a reasonable solution of the traffic flow optimization problem.
However, the original optimization problem has many local minima.
We may sample the binary variables following the effective Hamiltonian while tuning the Lagrange multiplier ${\boldsymbol \nu}$.
Below, we compare (i) the deterministic way by using the expectation (\ref{exact}), (ii) sampling by classical way following the Gibbs-Boltzmann distribution, and (iii) sampling by the D-Wave 2000Q.
The deterministic way quickly converges to the local minima of the cost function.
In the context of statistical mechanics, the deterministic way corresponds to the level of the mean-field analysis.
In this sense, this is a crude way to find an approximate solution.
The following two methods are beyond the mean-field analysis level because sample fluctuation occurs.
We utilize the sampling by the classical way following the Gibbs-Boltzmann distribution and by the D-Wave 2000Q just for selecting the choice of the route $\mu$ for each car $i$ depending on the value of $h_{\mu,i}$ while the outputs satisfy the constraints.
The essential difference between two procedures is in the intermediate dynamics.
The sampling by the classical way is based on hopping between the feasible solutions satisfying the constraint.
In contrast, the sampling by the D-Wave 2000Q is driven by the quantum tunneling effect.
The difference between two of the sampling methods appears as the performance of the resulting solutions.
The latter method leads to a slightly better solution than the former one as far as our observations in this problem setting are concerned.

The number of cars is set to be $350$ and that of the candidate routes is $3$ for each car.
The candidate routes are extracted from the actual maps.
When we straightforwardly implement the optimization problem, the system contains $1050$ spins and fully connected interactions, which is not directly solved by the D-Wave 2000Q.
As a reference, we put the solution from the Fujitsu digital annealer because the original optimization problem is difficult to implement directly on the D-Wave 2000Q.
We tune $\lambda$ to attain the best solution from the Fujistu digital annealer.

\begin{figure}
\begin{center}
\includegraphics[width = 1.0\textwidth]{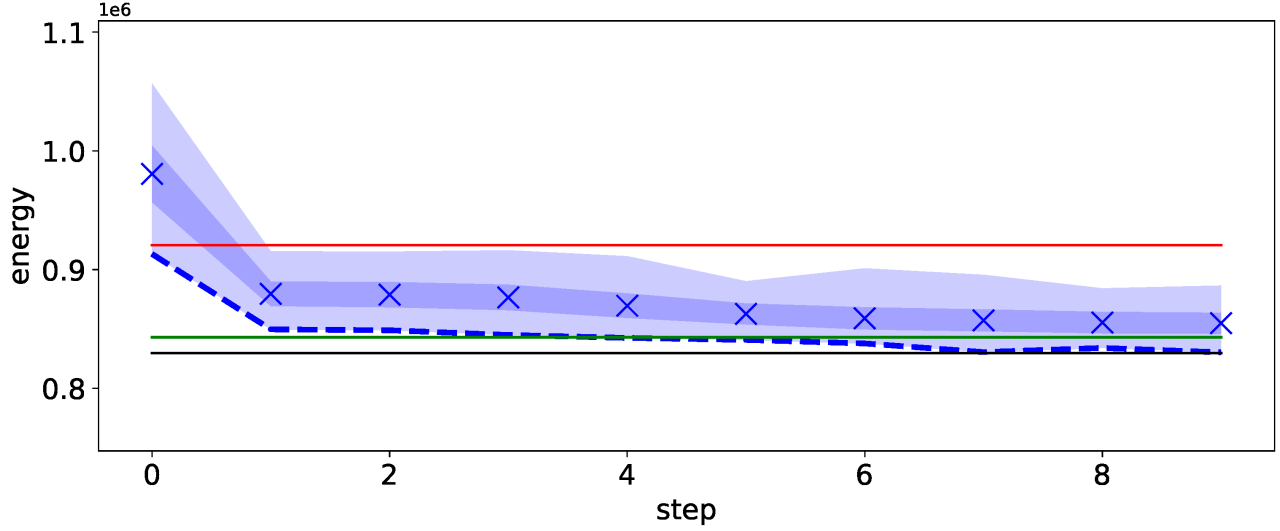}
\caption{Energy at each step in our method for optimizing the traffic flow at the Sendai city.
The same symbols are used as in Fig. \ref{np_ene} except for the lines.
In this figure, the red line denotes the result obtained by the deterministic way after a few steps, the green one is the minimum value by the sampling in the classical way during $10$ steps, and the black one represents the reference result attained by direct manipulation of the Fujitsu digital annealer.
}
\label{tfopt_ene}
\end{center}
\end{figure}

As shown in Fig. \ref{tfopt_ene}, we obtain the lower-energy solutions using the D-Wave 2000Q in comparison to the deterministic way and sampling by the classical way.
The results shown in Fig. \ref{tfopt_ene} satisfy the constraints for selecting the single route for each car because we do not apply our method in reduction of the quadratic term representing the constraints in this case.
To find solutions satisfying the constraints by use of our method, we need longer time to attain the feasible solutions.

\begin{figure}
\begin{center}
\includegraphics[width = 1.0\textwidth]{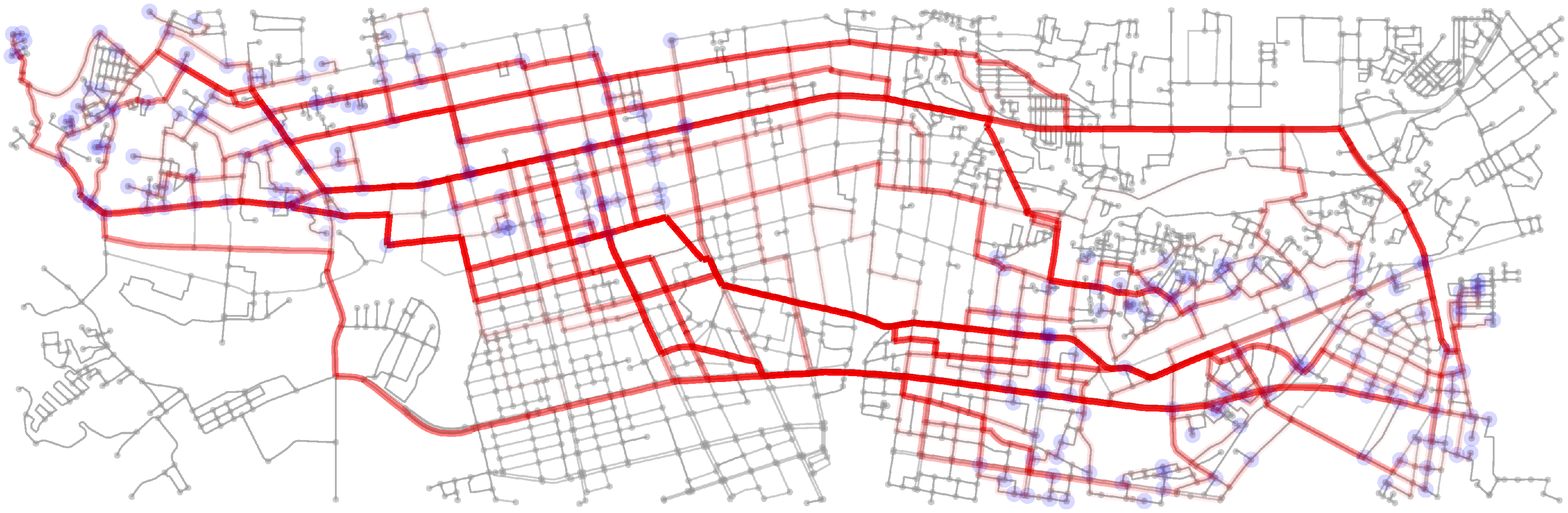}
\includegraphics[width = 1.0\textwidth]{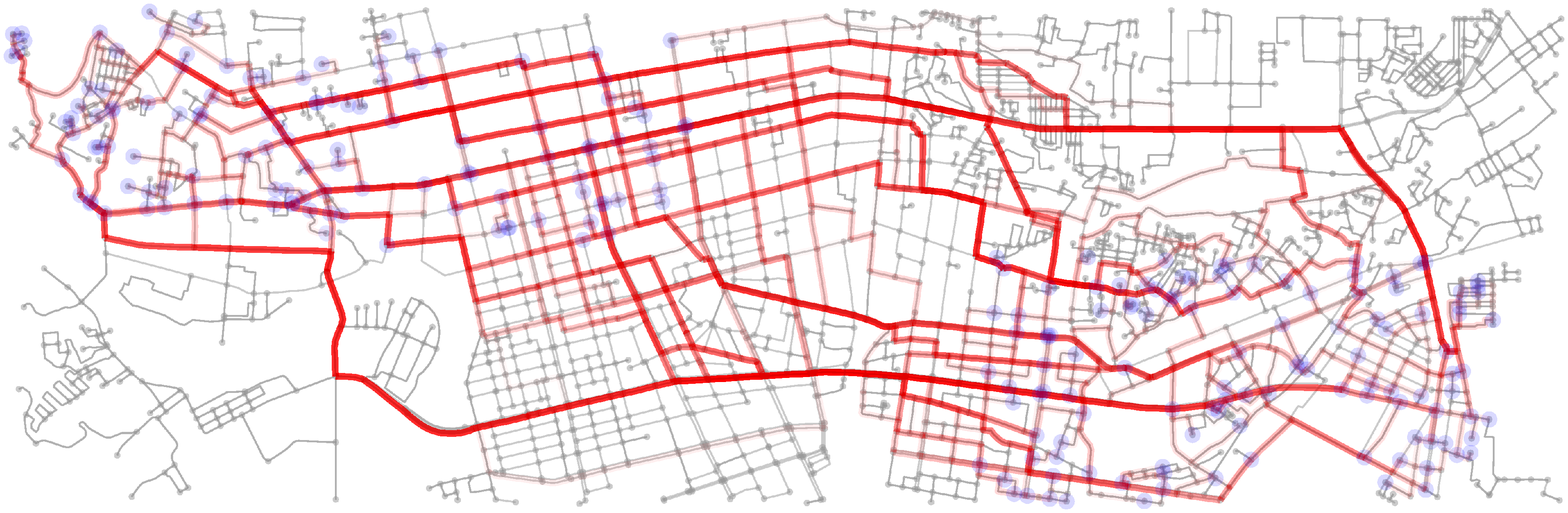}
\caption{Results in Sendai city obtained by (Upper panel) the shortest-path policy and (lower panel) our method.
The color strength of the red color on each road denoted by the edges represents the number of cars passing it.
The blue points are the starting and destination points.
We test our method around the point with $38^\circ 28' $ north latitude and $140^\circ 92'$ west longitude.
}
\label{tfopt_map}
\end{center}
\end{figure}

We test our method for the traffic flow optimization in Sendai city.
Sendai city and nearby areas suffer from disaster by Tsunamis after big earthquakes in $2011$.
The optimal solution provides the appropriate information for evacuation avoiding traffic jam.
The attained solutions are plotted in Fig. \ref{tfopt_map}.
For comparison, we have also plotted the shortest-path policy, in which each car selects the shortest path between the starting and destination points.
As a reference, the resulting cost function is given as $920697$ by the deterministic way, $848671$ sampling by the classical way, and $830309$ by our method with the D-Wave 2000Q, while the shortest path policy results in $1050159$.

\begin{figure}
\begin{center}
\includegraphics[width = 1.0\textwidth]{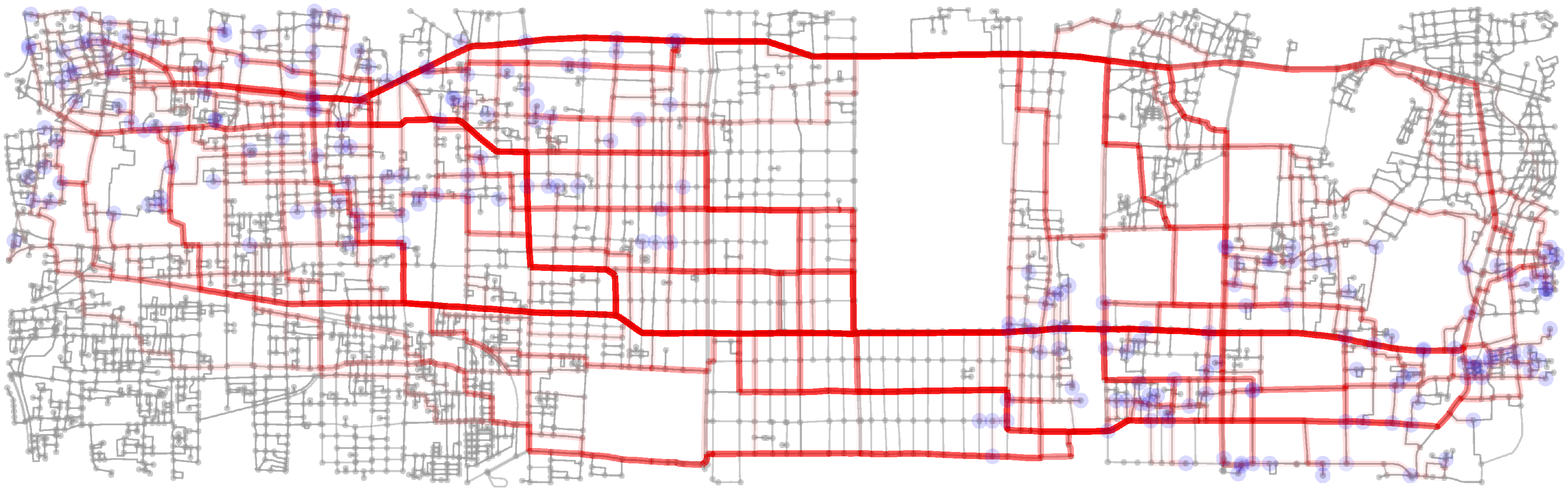}
\includegraphics[width = 1.0\textwidth]{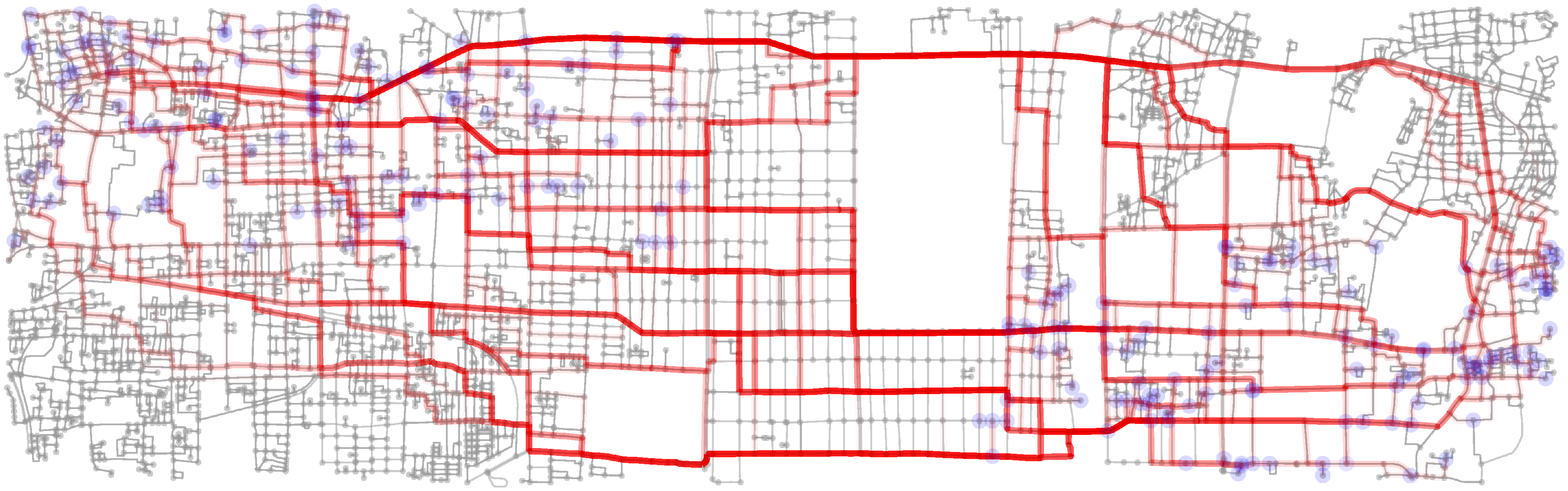}
\caption{Results in Kyoto city obtained by the (Upper panel) shortest-path policy and (lower panel) our method.
The same symbols and lines are used in Fig. \ref{tfopt_map}.
We tested our method around the point with $35^\circ 03' $ north latitude and $135^\circ 80'$ west longitude.
}
\label{tfopt_map2}
\end{center}
\end{figure}

In addition, we also tested our method in Kyoto city as shown in Fig. \ref{tfopt_map2}.
In this case, we attained the cost function to be $1602847$ by the deterministic way, $1288513$ sampling by the classical way, and $1284577$ by our method with the D-Wave 2000Q, while the shortest path policy results in $1782220$.

As demonstrated above, we solved the traffic-flow optimization problem exceeding the directly embeddable size on the D-Wave 2000Q.
The precision of the results is at essentially the same level as that of the Fujitsu digital annealer, which can directly solve the optimization problem with a large number of binary variables.

\begin{figure}
\begin{center}
\begin{tabular}{c}
\begin{minipage}{0.5\textwidth}
\includegraphics[width =\textwidth]{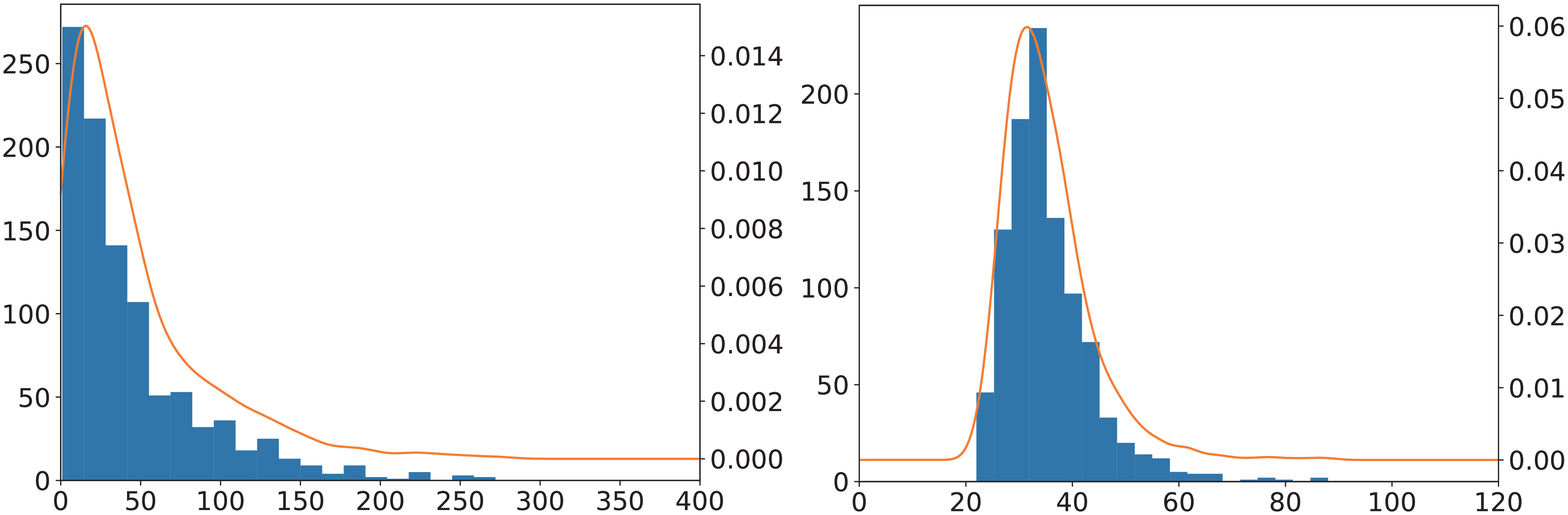}
\end{minipage}
\begin{minipage}{0.5\textwidth}
\includegraphics[width =\textwidth]{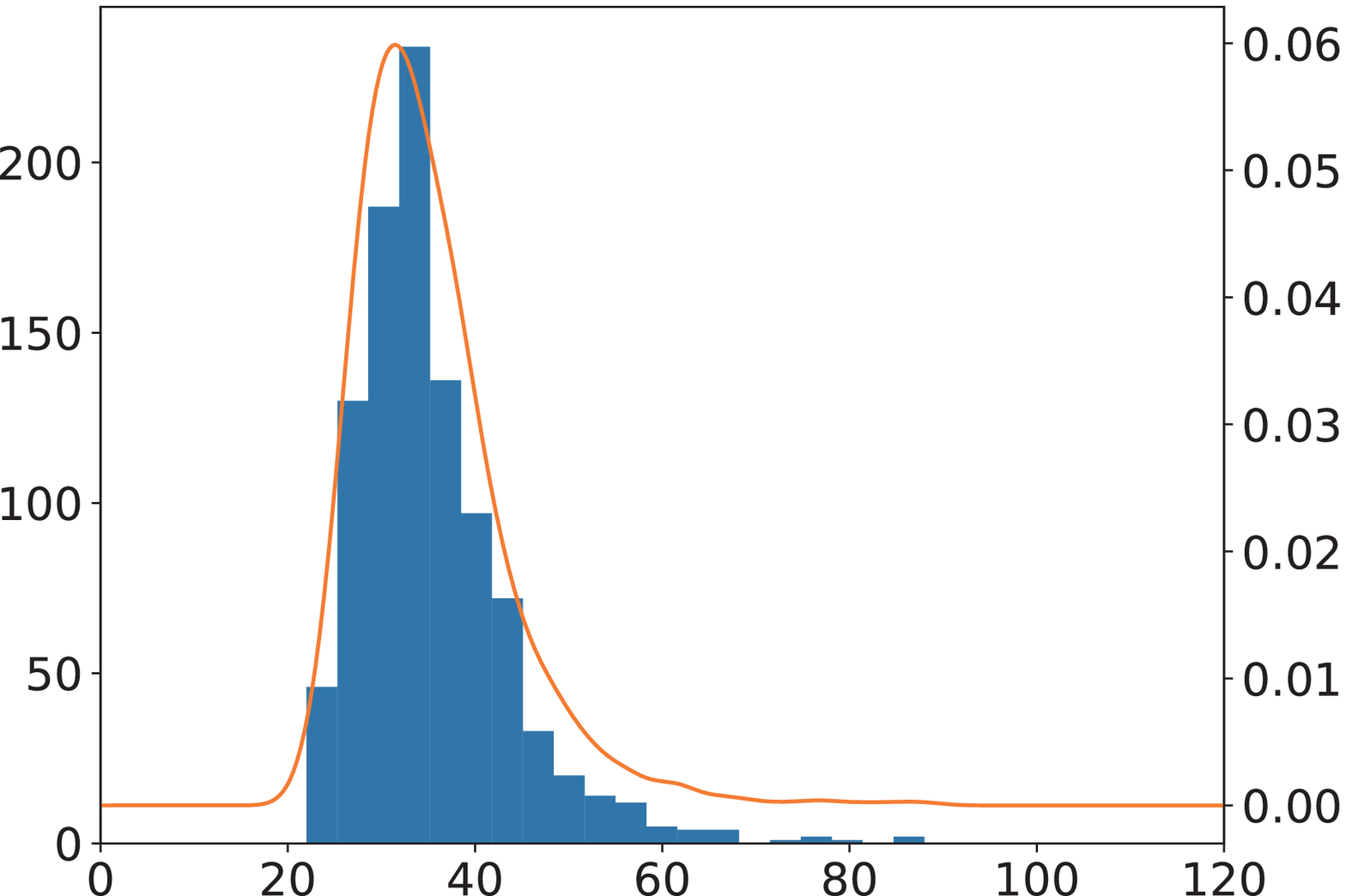}
\end{minipage}
\end{tabular}
\caption{Histogram of number of iterations for the cases to solve Eqs. (\ref{NP}) (left) and (\ref{CDMA}) (right one).
The horizontal axis denotes the number of iterations to attain the ground state.
The left vertical axis stands for occurrences of the number of steps in $1000$ runs and the right one represents the ratio.
The curves are attained by the Kernel density estimation as a guide for eyes. 
}
\label{steps}
\end{center}
\end{figure}

To investigate how many steps our method takes typically, we run it in the case of the number partition problems (\ref{NP}), which is the inference of the $N$-dimensional input (\ref{CDMA}) in $1000$ times.
Because we choose these examples, we know the ground state a priori.
As shown in Fig. \ref{steps}, all the cases converge to the ground state using our method and take several dozens of typical iterations.

\begin{figure}
\begin{center}
\begin{tabular}{c}
\begin{minipage}{0.5\textwidth}
\includegraphics[width = \textwidth]{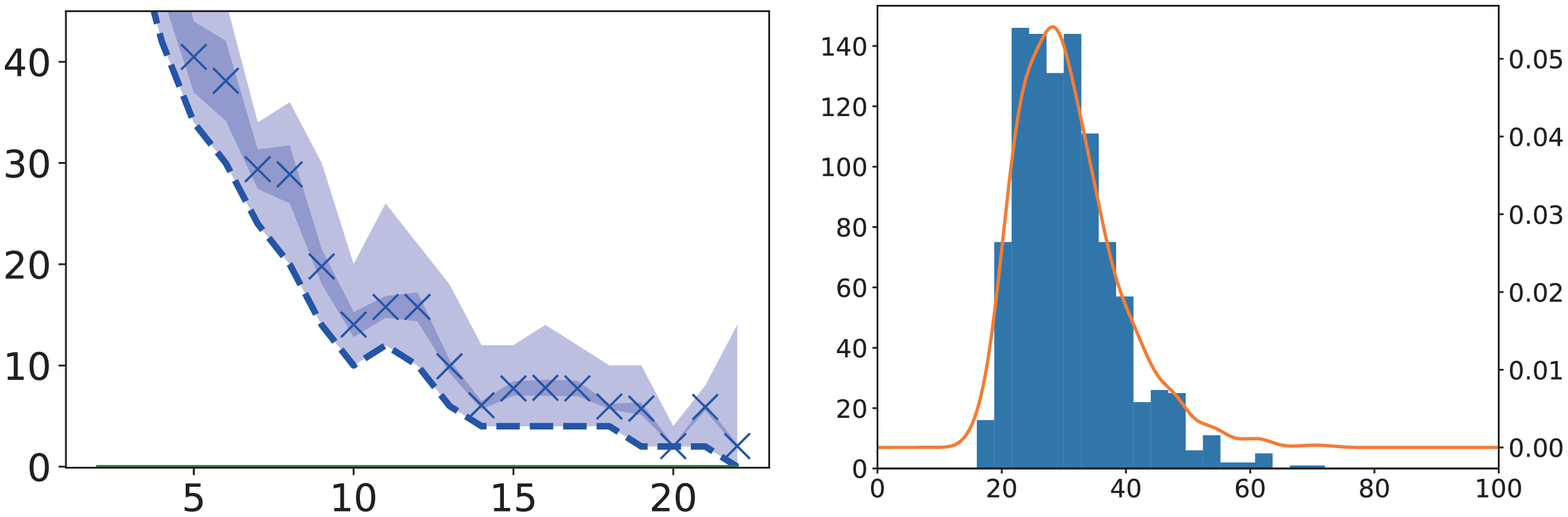}
\end{minipage}
\begin{minipage}{0.5\textwidth}
\includegraphics[width = \textwidth]{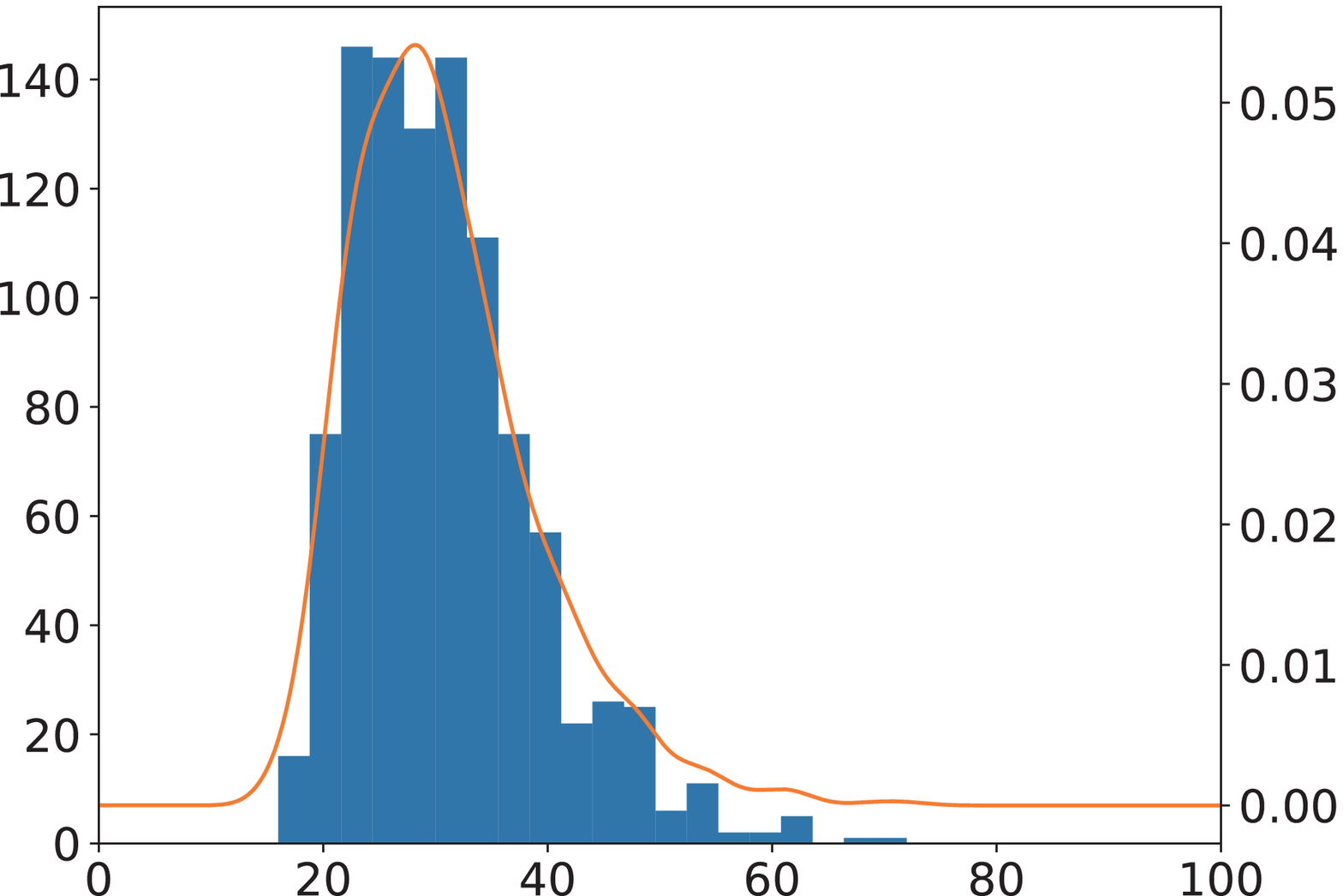}
\end{minipage}
\end{tabular}
\caption{
Summation of the second and third terms for each iteration (left) and histogram of the number of iterations in  solving Eq. (\ref{double_eff}) (right).
The same symbols are used in Figs. (\ref{cdma_ene}) and (\ref{steps}).
In the left panel, the vertical axis denotes the summation of the second and third terms in Eq. (\ref{double_con}) and the horizontal one represents the step.
}
\label{double_con}
\end{center}
\end{figure}

As the last example, we take a simple problem with double constraints, as is often seen in several practical optimization problems.
The original cost function is written as 
\begin{equation}
f({\bf q}) = \sum_{i,t}^N h_{it} q_{it} + \frac{\lambda}{2}\sum_{t=1}^L \left( \sum_{i=1}^L q_{it} - 1\right)^2+ \frac{\lambda}{2}\sum_{i=1}^L \left( \sum_{t=1}^L q_{it} - 1\right)^2, \label{double}
\end{equation}
where $h_{it}$ is the randomly generated values, $L$ is the linear size of the system, and the number of spins is $N=L\times L$.
This is the simplified version of the double-constraint problems as the traveling salesman problem.
In the traveling salesman problem, an agent moves to each city $i$ at each time $t$ only once.
To satisfy the rule, the cost function the double constraints as in the second and third terms as in Eq. (\ref{double}).
If we naively use the D-Wave 2000Q to solve this problem, the number of spins is limited to $N=64$ and thus the number of cities to $L=8$.
We instead consider the random-field Ising model with the double constraint as in Eq. (\ref{double}) to confirm the advantage of our method for satisfying such hard constraints.
We implement the following effective Hamiltonian and then deal with the number of spins, which drastically increases up to $L=45$, namely $N=2025$, as
\begin{equation}
H({\bf q},{\boldsymbol \nu}) = \sum_{\mu,i}h_{i,t} q_{i,t} + \sum_{i,t}\nu_{i,t} q_{i,t}.\label{double_eff}
\end{equation}
As in Fig. \ref{double_con}, we test our method to find the ground  state of the original cost  function (\ref{double}).

\section*{Summary}
We propose a technique to change various optimization problems with squared terms into those only with linear terms using the Hubbard-Stratonovich transformation.
The squared terms hamper efficient computation when special-purpose hardware, such as D-Wave 2000Q, is used to solve the optimization problems.
Our method mitigates the difficulty in dealing with the squared terms.
Instead of direct manipulation, we iteratively solve the optimization problem with linear terms and nontrivial terms.
We take various examples to test our methods.
The first one is to select $K$ variables under the random field, the second one is the number partition problem, the third one is to solve the linear equations, and the forth one is to reconstruct the structured data.
These are the optimization problems to find the feasible solutions satisfying the constraints.
Although our method can attain feasible solutions, it takes a long time to converge to them because a number of Lagrange multipliers need to be tuned.
In this sense, the application of our method is very important.

Apart from the previous four examples, the fifth one is the application of our method to give a lower-energy solution satisfying the constraints.
A part of the original optimization problem is reduced to the linear term, which becomes the local field.
We do not consider the constraints by our method in this case because it is easy to satisfy them under only the local field.
Then, our method leads to the feasible solution with lower energy.
To attain feasible solutions, we propose three of the methods.
The first one is the deterministic way to find the local minima, the second one is sampling by the classical way while jumping between feasible solutions.
The third one is the sampling by the D-Wave 2000Q for the binary variables, which do not necessarily satisfy the constraints.
In this sense, the range of the search for the optimal solutions is considered to be wide.
Thus, the third method is to efficiently find the better solution than the first and second ones.

In addition, these types of applications is a generalization of the straightforward application of our approach for the target Hamiltonian without any constraints, where we set $f_0({\bf q}) = 0$ and ${\bf C}={\bf 0}$, is written as
\begin{equation}
f({\bf q}) = {\bf q}^{\rm T}A{\bf q} \to {\boldsymbol \nu}^{\rm T}\sqrt{\Lambda} U {\bf q},
\end{equation}
where $A$ is a QUBO matrix, $\Lambda$ is the diagonal matrix including the eigenvalues $\lambda_k~(k=1,2,\cdots,N)$ of $A$, and $U$ is the orthogonal matrix diagonalizing $Q$.
Then $F_k({\bf q}) = {\bf u}^{\rm T}_k {\bf q}$.
In this case, the saddle-point equation is 
\begin{equation}
{\bf h} = A\left\langle {\bf q}\right\rangle_{\bf q},
\end{equation}
where ${\bf h} = {\boldsymbol \nu}_k^{\rm T}{\bf u}_k$ and
\begin{equation}
Q({\bf q}) = \frac{1}{Z({\boldsymbol \nu})}\exp\left( \beta {\bf h}^{\rm T}{\bf q} \right).
\end{equation}
Then the Taylor expansion of the Gibbs free energy $G({\bf 0})$ with respect to $A$ leads to the Plefka expansion.
The expansion up to the second order leads to saddle-point equation corresponding to the TAP equation.
In this sense, our approach is a generalization of the mean-field analysis.

In general, we may solve the optimization problem by changing the Lagrange multipliers iteratively.
On the D-Wave 2000Q, the fully connected interactions can be dealt with up to $64$ binary variables.
However, using our method, we can solve the QUBO including the Sherrington-Kirkpatrick model, which is a typical problem in spin glass theory, up to $2048$.
The sampling depending on the local field is an easy task.
However, the sampling with changing value of the local fields depending on the Lagrange multipliers has a history, and it crucially affects the performance of the resulting solutions.
We need another ingredient to improve the effect of the history of our method as proposed in the TAP equation to more efficiently solve the Ising spin-glass problem beyond the naive mean-field theory.
Possibly, the quantum tunneling effect might remove the effect of the history.
As far as our experience is concerned, we can find better solutions from the D-Wave 2000Q than from the classical way of sampling.
This will be detailed in a future study.

As pointed out in the previous section, the performance of our method is strongly dependent on the form of $f_0({\bf q})$.
All the cases tested in the present study have the simple forms of $f_0({\bf q})=0$ or linear combinations.
In the traveling salesman problem, the interactions $f_0({\bf q})= \sum_{t}\sum_{i,j}d_{i,j}q_{i,t}q_{j,t+1}$, where $d_{ij}$ is the distance between different cities $i$ and $j$.
Because $d_{ij} > 0$, the Griffiths inequality might not  hold in this case.
In other words, $\left\langle F_{k}({\bf q})\right\rangle_{\bf q}$ is not necessarily a monotonic increasing function against ${\boldsymbol \nu}$.
Therefore, the current version of our method might not be capable to efficiently lead to the ground state for the typical hard optimization problems.

We again emphasize that the original optimization problem solved in our study, which has fully connected interactions, cannot be embedded on the D-Wave 2000Q.
In this sense, our method makes a step to go ahead for more difficult tasks using the D-Wave 2000Q by reduction of the squared terms generating the fully connected interactions.
We actually reveal not only the potential of D-Wave 2000Q, but also CMOS annealing chip.
They do not suffer from the embedding of the optimization problem on the sparse graph due to the limitation of each piece of hardware.
In addition, our method makes it possible to deal with the four-body interaction.
By reducing the four-body interactions to the squared terms of the two-body interactions via diagonalization, we can obtain an effective two-body interacting system.
In this sense, our method reveals the capability to solve a wide range of Ising models by using the special-purpose hardware.
In addition, our method does not stick to the case to solve the optimization problem.
Because our technique is based on statistical mechanics, we utilize our method to perform efficient sampling at low temperatures.
We can find the hidden potential of the special-purpose hardware not only for solving the optimization problem but also for Boltzmann machine learning.

\bibliography{QApotential_ver2}
\section*{Acknowledgements}
The authors would like to thank Masamichi J. Miyama, Shuntaro Okada, Shunta Arai, and Shu Tanaka for the fruitful discussions. 
The present work was financially supported by JSPS KAKENHI Grant No. 19H01095, and Next Generation High-Performance Computing Infrastructures and Applications R\&D Program by MEXT.
The research was partially supported by National Institute of Advanced Industrial Science and Technology.
We utilized the Fujitsu digital annealer by courtesy of Fujitsu Limited.

\section*{Author contributions statement}
M.O. did the experiment, analyzed all the results, and wrote the manuscript.

\section*{Additional information}
{\bf Competing Interests}: The author declares that he has no competing interests.

\end{document}